\documentclass[prb,twocolumn,showpacs,preprintnumbers,amsmath,amssymb]{revtex4}
\usepackage{amsfonts}
\usepackage{graphicx}
\usepackage{dcolumn}
\usepackage{bm}

\renewcommand{\theequation}{\arabic{section}.\arabic{equation}}
\makeatletter\@addtoreset{equation}{section} \makeatother

\newcommand{\ind}[1]{\textrm{\footnotesize #1}}
\newcommand{\com}[1]{}

\begin{document}

\title{Exchange Interaction, Entanglement and Quantum Noise \\ Due to a Thermal Bosonic Field}

\author{Dmitry Solenov,\footnote{E-mail: Solenov@clarkson.edu}
Denis Tolkunov,\footnote{E-mail: Tolkunov@clarkson.edu} \ and\
Vladimir Privman\footnote{E-mail: Privman@clarkson.edu} }
\affiliation{Department of Physics, Clarkson University, Potsdam,
New York 13699--5820}


\begin{abstract}
We analyze the indirect exchange interaction between two two-state
systems, e.g., spins $1/2$, subject to a common finite-temperature
environment modeled by bosonic modes. The environmental modes,
e.g., phonons or cavity photons, are also a source of quantum
noise. We analyze the coherent vs\ noise-induced features of the
two-spin dynamics and predict that for low enough temperatures the
induced interaction is coherent over time scales sufficient to
create entanglement. A nonperturbative approach is utilized to
obtain an exact solution for the onset of the induced interaction,
whereas for large times, a Markovian scheme is used. We identify
the time scales for which the spins develop entanglement for
various spatial separations. For large enough times, the initially
created entanglement is erased by quantum noise. Estimates for the
interaction and the level of quantum noise for localized impurity
electron spins in Si-Ge type semiconductors are given.
\end{abstract}


\maketitle
\section{Introduction} \label{Section1}
The idea that exchange of fermionic or bosonic excitations can
lead to physical interactions in solid state is not
new.\cite{Mahan} Recently, such induced interactions have received
attention due to the possibility to experimentally observe quantum
dynamics in nanoscale devices.\cite{Jiang, Jiang2, Craig,
Elzerman, Koppens, Petta, MMJ, PF} In this work we explore the
dynamics of two qubits (two two-state quantum systems), e.g.,
electron spins $1/2$, placed a distance $\mathbf{d}$ apart, as
they are entangled by common thermalized bosonic environment
without direct spatial electron wave function overlap. At the same
time quantum noise originating from the same bosonic field (e.g.,
phonons) ultimately erases the generated entanglement for large
enought times. We demonstrate that the indirect exchange
interaction induced by the bosonic thermal field \cite{STPs} can,
in some cases, be comparable with other qubit-qubit couplings.

Extensive studies have been
reported\cite{Braun,Eberly1,Eberly2,Tolkunov1} of the decay of
quantum correlations between qubits subject to individual (local)
or common environmental noise. In the presence of quantum noise,
entanglement was shown to decay very fast and, in some cases,
vanish at finite times.\cite{Eberly1,Eberly2} On the other hand,
the idea that common bosonic as well as fermionic environment is
able to entangle the qubits has also been
advanced.\cite{Braun,STPs,RKKY,Bychkov,PVK,MPV,MPG,Piermarocchi,Porras,Mozyrsky,Rikitake}
For fermionic environment, this effect has been attributed to
Rudermann-Kittel-Kasuya-Yosida (RKKY) type
interactions\cite{RKKY,PVK,MPV,MPG,Piermarocchi,Mozyrsky,Rikitake}
and it has recently been experimentally demonstrated for two-qubit
semiconductor nanostructures.\cite{Craig,Elzerman}

In this work, we investigate the two competing physical effects of
a thermalized bosonic environment (bath) in which two qubits are
immersed. Specifically, the induced interaction, which is
effectively a zero-temperature effect, and the quantum noise,
originating from the same bath modes, are derived within a uniform
treatment. We study the dependence of the induced coherent vs\
noise (decoherence) effects on the parameters of the bath modes,
the qubit system, and their coupling, as well as on the geometry.
Specific applications are given for spins interacting with phonons
in semiconductors.

We assume that the spins are identically coupled with the modes of
a thermalized bosonic bath, described by
$H_B=\sum_{\mathbf{k},\mathbf{\xi}}\omega
_{\mathbf{k},\mathbf{\xi}}a_{\mathbf{
k},\mathbf{\xi}}^{\dagger}a_{\mathbf{k},\mathbf{\xi}}
$
, where we introduced the polarization index $\mathbf{\xi}$ and
set $\hbar=1$. With qubits represented as spins $1/2$, the
external magnetic field is introduced in the qubit Hamiltonian
\begin{equation}
H_S=\Delta (\sigma_z^1
+ \sigma_z^2 )/2 ,
\end{equation}
as the energy gap $\Delta$ between the up and down states for
spins 1 and 2, with the spins labeled by the superscripts. A
natural example of such a system are spins of two localized
electrons interacting via lattice vibrations (phonons) by means of
the spin-orbit interaction.\cite{Mahan,Hasegawa,Roth,SO-Winkler}
Another example is provided by atoms or ions in a cavity, used as
two-state systems interacting with photons. For each type of
phonon/photon, the interaction will be
assumed\cite{Mahan,SO-Winkler,Leggett,VKampen} linear in the
bosonic variables. The spin-boson coupling for two spins $j=1,2$
has the form
\begin{equation}\label{eq:S1:H_SB}
H_{SB}=\sum\limits_{j=1,2}\sum\limits_{m=x,y,z}\sigma^j_m X^j_m,
\end{equation}
where $\sigma^j_m$ are the standard Pauli matrices and
\begin{equation}\label{eq:S1:X_jm}
X^j_m=\sum_{\mathbf{k},\mathbf{\xi}}g^m_{\mathbf{k},\mathbf{\xi}}e^{i\mathbf{k}\cdot\mathbf{r}_j}
\left(a_{\mathbf{k},\mathbf{\xi}}+a_{-\mathbf{k},\mathbf{\xi}}^{\dagger}\right).
\end{equation}
\noindent{}Here, as before, the index $\xi$ accounts for the
polarization and $\mathbf{r}_j$ is the position of $j$th spin. The
overall system is described by the Hamiltonian
$H=H_S+H_B+H_{SB}$.
Our emphasis
will be on comparing the relative importance of the coherent
vs.\ noise
effects of a given bosonic bath in the two-qubit dynamics. We do
not include other possible two-qubit interactions in such
comparative calculation of dynamical quantities.

For the analysis of the induced exchange interaction and quantum
noise for most time scales of relevance, it is appropriate to use
the Markovian approach\cite{VKampen,Louisell,Blum} that assumes
instantaneous rethermalization of the bath modes. In
Section~\ref{Section2}, we derive in a unified formulation the
bath-induced spin-spin interaction and noise terms in the
dynamical equation for the spin density matrix. The onset of the
interaction for very short times is also investigated within an
exactly solvable model presented in Sections~\ref{Section3} and
\ref{Section4}. Specifically, in Section~\ref{Section4}, we
discuss the onset and development of the interaction Hamiltonian,
as well as the density matrix. It is shown that the initially
unentangled spins can develop entanglement. We find that the
degree of the entanglement and the time scale of its ultimate
erasure due to noise can be controlled by varying several
parameters, as further discussed for various bath types in
Section~\ref{Section5}. Estimates for the induced coherent
interaction in Si-Ge type semiconductors are presented in
Section~\ref{Section6}. The coherent interaction induced by
phonons is, expectedly, quite weak. However, we find that in
strong magnetic fields it can become comparable with the
dipole-dipole coupling.

\section{Coherent Interaction and Quantum Noise Induced by Thermalized Bosonic
Field} \label{Section2}
In this section we present the expressions for the induced
interaction and also for the noise effects due to the bosonic
environment, calculated perturbatively to the second order in the
spin-boson interaction, and with the assumption that the
environment is constantly reset to thermal. Specific applications
and examples are considered towards the end of this section, as
well as in Sections~\ref{Section5} and \ref{Section6}.

The dynamics of the system can be described by the equation for
the density matrix
\begin{equation}
i\dot \rho (t) = [H,\rho (t)] .
\end{equation}
In order to trace over the bath variables, we carry out  the
second-order perturbative expansion. This dynamical description is
supplemented by the Markovain
assumption\cite{Leggett,VKampen,Louisell,Blum} of resetting the
bath to thermal equilibrium, at temperature $T$, after each
infinitesimal time step, as well as at time $t=0$, thereby
decoupling the qubit system from the
environment.\cite{Leggett,VKampen} This is a physical assumption
appropriate for all but the shortest time scales of the system
dynamics.\cite{Privman,Tolkunov2,Solenov} It can also be viewed as
a means to phenomenologically account in part for the
randomization of the bath modes due to their interactions with
each other (anharmonicity) in real systems. This leads to the
master equation for the reduced density matrix of the qubits $\rho
_S (t) = Tr_B \rho (t) $,
\begin{eqnarray}\label{eq:S2:MME}
i\dot \rho _S (t) &=& [H_S ,\rho _S (t)]
\\ \nonumber
&-& i \int\limits_0^\infty {dt'Tr_B [H_{SB} ,[H_{SB} (t' - t),\rho
_B \rho _S (t)]]} ,
\end{eqnarray}
where $H_{SB} (\tau)=e^{i(H_B+H_S)\tau}H_{SB}e^{-i(H_B+H_S)\tau}$,
$\rho_B=\mathrm{Z}^{-1}\prod_k e^{-\omega_k a_k^\dag a_k/k_BT}$,
and $\mathrm{Z}=1/\prod_k (1-e^{-\omega_k/k_B T})$ is the
partition function. Analyzing the structure of the integrand in
Eq.(\ref{eq:S2:MME}), one can obtain the equation with explicitly
separated coherent and noise contributions, see
Appendix~\ref{SecA},
\begin{equation}\label{eq:S2:MRateEq}
i\dot \rho _S (t) = [H_\textrm{eff} ,\rho _S (t)] + i{\hat M}\rho
_S (t).
\end{equation}
Here the effective coherent Hamiltonian $H_\textrm{eff}$ is
\begin{eqnarray}\nonumber
H_\textrm{eff}\!\!\!&=&\!\!\!  H_S  +\!\!\! \sum\limits_{m=x,y,z} \!\!\! {2\chi _c^m
({\mathbf{d}})\sigma _m^1 \sigma _m^2 }  - \chi _s^x
({\mathbf{d}})\left( {\sigma _x^1 \sigma _y^2  + \sigma _x^2
\sigma _y^1 } \right)
\\\nonumber \!\!\! &+&\!\!\! \chi _s^y ({\mathbf{d}})\left( {\sigma _y^1 \sigma _x^2  +
\sigma _y^2 \sigma _x^1 } \right) -
\left[\eta_s^x(0)+\eta_s^y(0)\right]
\left({\sigma_z^1+\sigma_z^2}\right). \\\label{eq:S2:H-eff}
\end{eqnarray}
The expressions for the amplitudes $\chi_c^m(\mathbf{d})$,
$\chi_s^m(\mathbf{d})$, $\eta_c^m(\mathbf{d})$, and
$\eta_s^m(\mathbf{d})$ will be given shortly. The first three
terms following $H_S$ constitute the interaction between the two
spins. We will argue below that the leading induced exchange
interaction is given by the first added term, proportional to
$\chi_c^m(\mathbf{d})$. The last term corrects the energy gap for
each qubit, representing their Lamb shifts.

The explicit expression for the noise term is very cumbersome. It
can be represented concisely by introducing the noise
superoperator $\hat M$, which involves single-qubit contributions,
which are usually dominant, as well as two-qubits terms
\begin{equation}\label{eq:S2:M-sums}
{\hat M} = \sum\limits_{m,i} \Big[ {\hat M}_m^i (0)
+\sum_{j \ne i} {\hat M}_m^{ij} (\mathbf{d} ) \Big] ,
\end{equation}
where the summations are over the components, $m=x,y,z$, and the
qubits, $i,j=1,2$. The quantities entering Eq.(\ref{eq:S2:M-sums})
can be written in terms of the amplitudes $\chi_s^m(\mathbf{d})$,
$\eta_c^m(\mathbf{d})$, and $\eta_s^m(\mathbf{d})$, in a compact
form, by introducing the superoperators ${\hat L}_a
(O_1)O_2=\{O_1,O_2\}$, ${\hat L}(O_1,O_2)O_3=O_1O_3O_2$, and
${\hat L}_\pm(O_1,O_2)={\hat L}(O_1,O_2)\pm {\hat L}(O_2,O_1)$,
\begin{eqnarray}\nonumber
{\hat M}_m^{ij}(\mathbf{d})&=&\eta_c^m(\mathbf{d}) \left[{2{\hat
L}(\sigma^i_m,\sigma^j_m) - {\hat L}_a(\sigma^i_m\sigma^j_m)}
\right]
\\ \label{eq:S2:M-cross}
&+& \eta_s^m(\mathbf{d})\left[{{\hat
L}_+(\sigma^i_m,\varsigma^j_{m})-{\hat L}_a(\sigma^i_m
\varsigma^j_{m} )}\right]
\\ \nonumber
&-&i\chi_s^m(\mathbf{d}){\hat L}_-(\sigma^i_m ,\varsigma^j_{m}),
\end{eqnarray}
where we defined $\varsigma^j_{m} = \frac{i}{2}[\sigma_z^j ,\sigma
_m^j]$ and
\begin{eqnarray}\nonumber
{\hat M}_m^j(0)&=&\eta _c^m (0)\left[{2{\hat L}(\sigma _m^j,\sigma
_m^j) -{\hat L}_a(\sigma _m^j \sigma _m^j )}\right]
\\ \label{eq:S2:M-local}
&+&\eta _s^m(0){\hat L}_+(\sigma_m^j ,\varsigma^j_{m})
\\ \nonumber
&-&i\chi _s^m (0)\left[{{\hat
L}_-(\sigma_m^j,\varsigma^j_{m})+{\hat L}_a(\sigma_m^j
\varsigma^j_{m})}\right].
\end{eqnarray}

The amplitudes in Eqs.(\ref{eq:S2:H-eff}, \ref{eq:S2:M-cross},
\ref{eq:S2:M-local}), calculated for the interaction defined in
Eqs.(\ref{eq:S1:H_SB}, \ref{eq:S1:X_jm}), are
\begin{equation}\label{eq:S2:he-c}
\chi _c^m ({\mathbf{d}}) =  - \sum\limits_\mathbf{\xi}
{\int\limits_{ - \infty }^\infty  {\frac{{Vd{\mathbf{k}}}}
{{\left( {2\pi } \right)^3 }}} } \left|
{g_{{\mathbf{k}},\mathbf{\xi}}^m } \right|^2 \frac{{\omega
_{{\mathbf{k}},\mathbf{\xi}} \cos \left( {{\mathbf{k}} \cdot
{\mathbf{d}}} \right)}} {{\omega _{{\mathbf{k}},\mathbf{\xi}}^2  -
\Delta ^2 (1 - \delta _{m,z} )}},
\end{equation}
\begin{eqnarray}\nonumber
\eta _c^m ({\mathbf{d}}) &=& \frac{\pi}{2}\sum\limits_\mathbf{\xi}
{\int\limits_{ - \infty }^\infty  {\frac{{Vd{\mathbf{k}}}}
{{\left( {2\pi } \right)^3 }}} } \left|
{g_{{\mathbf{k}},\mathbf{\xi}}^m } \right|^2 \coth \frac{{\omega
_{{\mathbf{k}},\mathbf{\xi}} }} {{2k_\textrm{B}T}}\cos \left(
{{\mathbf{k}} \cdot {\mathbf{d}}} \right)
\\ \label{eq:S2:eta-c}
&\times&\sum\limits_{q = \pm 1} {\delta \left( {\omega
_{{\mathbf{k}},\mathbf{\xi}} + (1 - \delta _{m,z} )q\Delta }
\right)},
\end{eqnarray}
and
\begin{eqnarray}\nonumber
\chi _s^m ({\mathbf{d}}) &=&  - \left( {1 - \delta _{m,z} }
\right)\sum\limits_\mathbf{\xi} {\int\limits_{ - \infty }^\infty
{\frac{{Vd{\mathbf{k}}}} {{\left( {2\pi } \right)^3 }}} } \left|
{g_{{\mathbf{k}},\mathbf{\xi}}^m } \right|^2 \cos \left(
{{\mathbf{k}} \cdot {\mathbf{d}}} \right)
\\ \label{eq:S2:he-s}
&\times&\sum\limits_{q =  \pm 1} {\frac{{\pi s}} {2}\delta
\left( {\omega _{{\mathbf{k}},\mathbf{\xi}}  + q\Delta } \right)},
\end{eqnarray}
\begin{eqnarray}\label{eq:S2:eta-s}
\eta _s^m ({\mathbf{d}}) &=& \left( {1 - \delta _{m,z} } \right)
\\ \nonumber
&\times&\sum\limits_\mathbf{\xi} {\int\limits_{ - \infty }^\infty
{\frac{{Vd{\mathbf{k}}}} {{\left( {2\pi } \right)^3 }}} } \left|
{g_{{\mathbf{k}},\mathbf{\xi}}^m } \right|^2 \coth \frac{{\omega
_{{\mathbf{k}},\mathbf{\xi}} }} {{2k_\textrm{B}T}}\frac{{\Delta
\cos \left( {{\mathbf{k}} \cdot {\mathbf{d}}} \right)}} {{\omega
_{{\mathbf{k}},\mathbf{\xi}}^2  - \Delta ^2 }}.
\end{eqnarray}
Here the principal values of integrals are assumed.

Note that $\chi _c^m ({\mathbf{d}})$ appears only in the induced
interaction Hamiltonian in Eq.(\ref{eq:S2:H-eff}), whereas $\eta
_c^m ({\mathbf{d}})$, $\chi _s^m ({\mathbf{d}})$, and $\eta _s^m
({\mathbf{d}})$ enter both the interaction and noise terms.
Therefore, in order to establish that the induced interaction can
be significant for some time scales, we have to demonstrate that
$\chi _c^m ({\mathbf{d}})$ can have a much larger magnitude than
the maximum of the magnitudes of $\eta _c^m ({\mathbf{d}})$, $\chi
_s^m ({\mathbf{d}})$, and $\eta _s^m ({\mathbf{d}})$. The third
and fourth terms in expression for the interaction
(\ref{eq:S2:H-eff}) are comparable to the noise and therefore have
no significant contribution to the coherent dynamics.

Because of the complexity of the expressions for the noise terms
within the present Markovian treatment, in this section we will
only compare the magnitudes of the coherent vs\ noise effects. In
the next section, we will discuss a different model for the noise
which will allow a more explicit investigation of the time
dependence.

For the rest of this section, we will consider an illustrative
one-dimensional (1D) example favored by recent
experiments,\cite{Craig,Elzerman} leaving the derivations for
higher dimensions to Sections~\ref{Section5} and \ref{Section6}. We comment that the 1D
geometry is also natural for certain ion-trap quantum-computing
schemes, in which ions in a chain are subject to Coulomb
interaction, developing a variety of phonon-mode lattice
vibrations.\cite{Porras,Marquet,Leibfried}

In 1D geometry we allow the phonons to propagate in a single
direction, along $\mathbf{d}$, so that ${\mathbf{k}} \cdot
{\mathbf{d}} = k\left| {\mathbf{d}} \right|$. Here, for
definiteness, we also assume the linear dispersion, $\omega _k  =
c_s k$, since the details of the dispersion relation for larger
frequencies usually have little effect on the decoherence
properties. Furthermore, we ignore the polarization,
$g_{{\mathbf{k}},\mathbf{\xi}}^m  \to g^m (\omega )$. Another
reason to focus on the low-frequency modes is that an additional
cutoff $\omega _c$ resulting from the localization of the electron
wave functions, typically much smaller than the Debye frequency,
will be present due to the factors $g^m (\omega )$. The induced
interaction and noise terms depend on the amplitudes $\chi _c^m
({\mathbf{d}})$, $\eta _c^m ({\mathbf{d}})$, $\chi _s^m
({\mathbf{d}})$, and $\eta _s^m ({\mathbf{d}})$, two of which can
be evaluated explicitly for the 1D case, because of the
$\delta$-functions in Eqs.(\ref{eq:S2:eta-c}, \ref{eq:S2:he-s}).
However, to derive an explicit expression for $\chi _c^m
({\mathbf{d}})$ and $\eta _s^m ({\mathbf{d}})$, one needs to
specify the $\omega$-dependence in $\left| {g^m (\omega )}
\right|^2$. For the sake of simplicity, in this section we
approximate $|g^m (\omega )|^2$ by a linear function with
superimposed exponential cutoff. For a constant 1D density of
states $\Upsilon (\omega ) = V/2\pi c_s$, this is the
Ohmic-dissipation condition,\cite{Leggett} i.e.,
\begin{equation}\label{eq:S2:DOS}
|g^m (\omega )|^2\Upsilon (\omega )=\alpha^m_n\omega^n
\exp(-\omega/\omega_c),
\end{equation}
with $n=1$ (the case when $n>1$ is considered in Section \ref{Section5}).

In most practical applications, we expect that $\Delta \ll
\omega_c$. With this assumption, we obtain
\begin{equation}\label{eq:S2:he-c-explicit-1D}
\chi _c^m ({\mathbf{d}}) = \frac{{\alpha _1^m \omega _c }} {{1 +
\left( {{{\omega _c \left| {\mathbf{d}} \right|} \mathord{\left/
 {\vphantom {{\omega _c \left| {\mathbf{d}} \right|} {c_s }}} \right.
 \kern-\nulldelimiterspace} {c_s }}} \right)^2 }},
\end{equation}
\begin{equation}\label{eq:S2:he-s-explicit-1D}
\chi _s^m ({\mathbf{d}}) = \alpha _1^m \omega _c \frac{\pi }
{2}\left( {1 - \delta _{m,z} } \right)\frac{\Delta } {{\omega _c
}}\cos \frac{{\Delta \left| {\mathbf{d}} \right|}} {{c_s }},
\end{equation}
and
\begin{equation}\label{eq:S2:eta-c-explicit-1D}
\eta _c^m ({\mathbf{d}}) = \alpha _1^m \omega _c \frac{\pi }
{2}\left( {1 - \delta _{m,z} } \right)\frac{\Delta } {{\omega _c
}}\coth \frac{\Delta } {{2k_\textrm{B}T}}\cos \frac{{\Delta \left|
{\mathbf{d}} \right|}} {{c_s }}.
\end{equation}
The expression for $\eta _s({\mathbf{d}})$ could not be obtained
in closed form. However, numerical estimates suggest that  $\eta
_s({\mathbf{d}})$ is comparable to $\eta _c^m ({\mathbf{d}})$. At
short spin separations $\mathbf{d}$, $\eta _s({\mathbf{d}})$ is
approximately bounded by $- \alpha_1^m \Delta
\ln\frac{\Delta}{\omega_c} \exp(-\frac{{\Delta \left| {\mathbf{d}}
\right|}} {{c_s }})$, while at lager distances it may be
approximated by $\alpha_1^m \Delta\frac{\pi}{2}\sin \frac{{\Delta
\left| {\mathbf{d}} \right|}} {{c_s }}
\coth\frac{\Delta}{2k_\textrm{B}T}$. The level of noise may be
estimated by considering the quantity
\begin{equation}\label{eq:S2:M-explicit-1D}
{\cal M} = \max_{|\mathbf{d}|} \left|{\eta _c^m(\mathbf{d}), \chi
_s^m(\mathbf{d}), \eta _s^m(\mathbf{d})} \right|.
\end{equation}

The interaction
Hamiltonian takes the form
\begin{equation}\label{eq:S2:H-int-1D}
H_{\operatorname{int} }  =  - \frac{2} {{1 + \left( {{{\omega _c
\left| {\mathbf{d}} \right|}/{c_s }}} \right)^2 }}
 \sum\limits_{m = x,y,z} {\alpha _1^m \omega _c \sigma _m^1 \sigma
 _m^2}.
\end{equation}
This induced interaction is temperature independent. It is
long-range and decays as a powerlaw for large $\mathbf{d}$. For
the super-Ohmic case, $n>1$, one obtains similar behavior, except
that the interaction decays as a higher negative power of
$\mathbf{d}$, as will be shown in Section~\ref{Section5}.

If the noise term were not present, the spin system would be
governed by the Hamiltonian $H_S  + H_{\textrm{int}}$. To be
specific, let us analyze the spectrum, for instance, for $\alpha
_1^x  = \alpha _1^y$, $\alpha _1^z  = 0$. The two-qubit states
consist of the singlet $(\left|{\uparrow\downarrow}\right\rangle-
\left|{\downarrow\uparrow}\right\rangle)/\sqrt{2}$ and the split
triplet $\left|{\uparrow\uparrow}\right\rangle$,
$(\left|{\uparrow\downarrow}\right\rangle+
\left|{\downarrow\uparrow}\right\rangle)/\sqrt{2}$, and
$\left|{\downarrow\downarrow}\right\rangle$, with energies
$E_2=-4\chi _c^x$, $E_0=-\Delta$, $E_1=4\chi _c^x$, $E_3=\Delta$,
respectively. The energy gap $|E_1-E_2|$ between the two entangled
states is defined by $4\chi _c^x$ ($=4\chi _c^y$). In the presence
of noise, the oscillatory, approximately coherent evolution of the
spins can be observed over several oscillation cycles provided
that $2\alpha_1^m \omega_c/[1 +(\omega_c |\mathbf{d}|/c_s)^2]>
{\cal M}$. The energy levels will acquire effective width due to
quantum noise, of order $\eta_c^m(\mathbf{d})$. This interplay
between the interaction and noise effects is further explored
within an exactly solvable model in the next section.
\begin{figure}
\includegraphics[width=7cm]{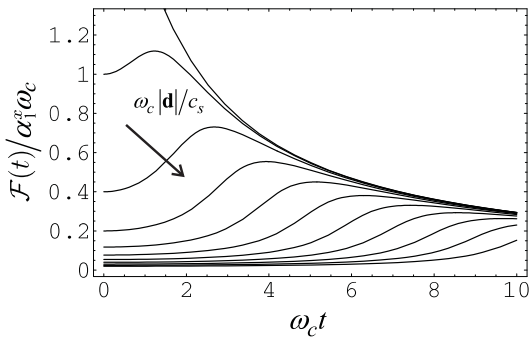}
\caption{Short-time correction to the induced exchange interaction
for the Ohmic case. The arrow shows the order of the curves for
increasing $\omega_c|\mathbf{d}|/c_s=0,1,...,10$.}\label{fig1}
\end{figure}

\section{Exactly Solvable Model} \label{Section3}

In this section, we consider a model appropriate for short
times,\cite{Privman,Tolkunov2,Solenov} which does not invoke the
Markovian assumption of the rethermalization of the bath modes.
This model is particularly suitable for investigating the onset
of the system's dynamics. While the noise effects are quantitatively
different in this model, the qualitative interplay of the coherent
and noise effects in the dynamics is the same as in the Markovian
approach. Furthermore, we will show that the
induced interaction is consistent with the one obtained within
Markovian approach in the previous section.

We point out that, due to the instantaneous rethermalization
assumption (resetting the density matrix of the bath to thermal),
in the Markovain formulation it was quite natural to assume that
the bath density matrix is also thermal at time $t=0$; the total
density matrix retained an uncorrelated-product form at all times.
In the context of studying the short-time dynamics, in this
section the choice of the initial condition must be addressed more
carefully. In quantum computing applications, the initially
factorized initial condition has been widely used for the
qubit-bath system\cite{Privman,Tolkunov2,Solenov}
\begin{equation}\label{eq:S3:IC}
\rho(0) = \rho_S(0)\rho_B(0).
\end{equation}
This choice allows comparison with the Markovian results, and is
usually needed in order to make the short-time approximation
schemes tractable,\cite{Solenov} specifically, it is necessary for
exact solvability of the model considered in this and the next
sections.

A somewhat more ``physical'' excuse for factorized initial
conditions has been the following. Quantum computation is carried
out over a sequence of time intervals during which various
operations are performed on individual qubits and on pairs of qubits.
These operations include control gates, measurement, and error
correction. It is usually assumed that these ``control''
functions, involving rather strong interactions with external
objects, as compared to interactions with sources of quantum
noise, erase the fragile entanglement with the bath modes that
qubits can develop before those time intervals when they are
``left alone'' to evolve under their internal (and bath induced)
interactions. Thus, for evaluating relative importance of the
quantum noise effects on the internal (and bath induced) qubit
dynamics, which is our goal here, we can assume that the state of
qubit-bath system is ``reset'' to uncorrelated at $t=0$.

It turns out that the resulting model is exactly solvable for the
Zeeman splitting $\Delta =0$, and provided that only a single
system operator enters the expression (\ref{eq:S1:H_SB}) for the
interaction. Here we take $\alpha _n^y=\alpha _n^z=0$, while
$\alpha _n^x \neq 0$. We derive the exact solution and demonstrate
the emergence of the interaction (\ref{eq:S2:H-int-1D}).

\begin{figure}
\includegraphics[width=8cm]{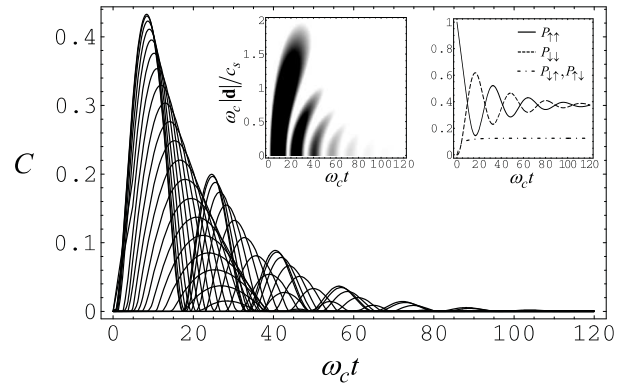}
\caption{Development of the concurrence as a function of time,
calculated with $\alpha^x_1=1/20$ and
$k_\textrm{B}T/\omega_c=1/20$. The curves correspond to various
spin-spin separations, as can be read off the left inset, which
shows the distribution of the concurrence in the
$|\mathbf{d}|$-$t$ plane. The right inset presents the dynamics of
the diagonal density matrix elements $P_{\uparrow\uparrow} \equiv
\langle \uparrow \uparrow | \rho_S | \uparrow \uparrow \rangle $,
etc., on the same time scale.}\label{fig2}
\end{figure}

With the above assumptions, one can utilize the bosonic operator
techniques\cite{Louisell} to obtain the reduced density matrix for
the system (\ref{eq:S1:H_SB}, \ref{eq:S1:X_jm}),
\begin{equation}\label{eq:S3:AdiabaticSolution}
\rho _S (t) = \sum\limits_{\lambda ,\lambda '} {P_\lambda  \rho _S
(0)P_{\lambda '} e^{\frak{L}_{\lambda \lambda '} (t)} },
\end{equation}
where the projection operator is defined as
$P_\lambda=\left|{\lambda_1\lambda_2}\right\rangle\left\langle
{\lambda _1\lambda _2}\right|$, and
$\left|{\lambda_j}\right\rangle$ are the eigenvectors of
$\sigma_x^j$. The exponent in Eq.(\ref{eq:S3:AdiabaticSolution})
consists of the real part, which leads to decay of off-diagonal
density-matrix elements resulting in decoherence,
\begin{eqnarray}\nonumber
  \textrm{Re} \frak{L}_{\lambda \lambda '}(t)&=&\!\!-\!\!\sum\limits_k {G_k (t,T)
  \left[ {\left( {\lambda '_1  - \lambda _1 } \right)^2  + \left({
  \lambda '_2  - \lambda _2 } \right)^2 } \right.}
\\ \label{eq:S3:DecoherenceFunction-ReL}
   &+&\!\!\! \left. {2\cos\! \left( {\frac{{\omega _k \left| {\mathbf{d}} \right|}}
{{c_s }}} \right)\!\left( {\lambda '_1  - \lambda _1 }
\right)\left( {\lambda '_2  - \lambda _2 } \right)} \right]
\end{eqnarray}
and the imaginary part, which describes the coherent evolution,
\begin{equation}\label{eq:S3:DecoherenceFunction-ImL}
\textrm{Im} \frak{L}_{\lambda \lambda '} (t) =  \sum\limits_k {C_k
(t)\cos \left( {\frac{{\omega _k \left| {\mathbf{d}} \right|}}
{{c_s }}} \right)\left( {\lambda _1 \lambda _2  - \lambda '_1
\lambda '_2 } \right)}.
\end{equation}
Here we defined the standard spectral
functions\cite{Leggett,Privman}
\begin{equation}\label{eq:S3:DecoherenceFunction-Gk}
G_k (t,T) = 2\frac{{\left| {g_k } \right|^2 }} {{\omega _k^2
}}\sin ^2 \left( \frac{{\omega _k t}} {{2}}\right) \coth \left(
{\omega _k \over 2k_B T}  \right)
\end{equation}
and
\begin{equation}\label{eq:S3:DecoherenceFunction-Ck}
C_k (t) = 2\frac{{\left| {g_k } \right|^2 }} {{\omega _k^2
}}\left( {\omega _k t - \sin \omega _k t} \right).
\end{equation}
Calculating the sums by converting them to integrals over the
bath-mode frequencies $\omega$ in
Eqs.(\ref{eq:S3:DecoherenceFunction-ReL}) and
(\ref{eq:S3:DecoherenceFunction-ImL}), assuming the Ohmic bath
$n=1$ for $T>0$ one obtains a linear in time $t$ large-time
behavior for both the temperature-dependent real part and for the
imaginary part. The coefficient for the former is $\sim\! kT$,
whereas for the latter it is $\sim\! \omega_c$. For super-Ohmic
models, $n>1$, the real part grows slower, as was also noted in
the literature.\cite{VKampen, Privman, PALMA}

Let us first analyze the effect that the imaginary part of
$\frak{L}_{\lambda \lambda '}(t)$ has on the evolution of the
reduced density matrix, since this contribution leads to the
induced interaction. If the real part were not present, i.e.,
omitting Eq.(\ref{eq:S3:DecoherenceFunction-ReL}) from
Eqs.(\ref{eq:S3:AdiabaticSolution}),
(\ref{eq:S3:DecoherenceFunction-ImL}), and
(\ref{eq:S3:DecoherenceFunction-Ck}), we would obtain the
evolution operator in the form $\exp[-iH_\textrm{int}t-iF(t)t]$.
The interaction $H_\textrm{int}$ comes from the first term in
Eq.(\ref{eq:S3:DecoherenceFunction-Ck}),
\begin{equation}\label{eq:S3:H-int}
H_\textrm{int}\! = \! - \frac{2\alpha _n^x \Gamma (n)c_s^n\omega
_c^n } {\left(c_s^2 + \omega _c^2 \left| {\mathbf{d}} \right|^2
\right)^{n/2} } \cos \left[{
 n\arctan \left( {\frac{{\omega _c \left| {\mathbf{d}} \right|}}
{{c_s }}} \right)} \right]\!\sigma _x^1 \sigma _x^2.
\end{equation}
This expression is the same as the results obtained within the
Markovian scheme, cf., Eqs.(\ref{eq:S2:H-int-1D}),
(\ref{eq:S2:he-c}), and Section~\ref{Section5}. The operator
$F(t)$ is given by
\begin{equation}\label{eq:S3:F(t)}
F(t) = 2\sigma _x^1 \sigma _x^2 \int\limits_0^\infty  {d\omega
\frac{{D(\omega )\left| {g(\omega )} \right|^2 }} {\omega
}\frac{{\sin \omega t}} {{\omega t}}\cos \left( \frac{\omega
|\mathbf{d}|}{c_s} \right)} .
\end{equation}
It commutes with $H_\textrm{int}$ and therefore could be viewed as
the initial time-dependent correction to the interaction. In fact,
it describes the onset of the induced coherent interaction; note
that $F(0)=-H_\textrm{int}$, but for large times $F(t) \sim
\alpha^x_n \omega_c^n/(\omega_c t)^n$. In Figure~\ref{fig1}, we
plot ${\cal F}(t)$, defined via $F(t)={\cal F}(t)\sigma_x^1
\sigma_x^2$, for the Ohmic case as a function of time for various
spin-spin separations.

Let us now explore the role of the decoherence term
(\ref{eq:S3:DecoherenceFunction-ReL}). In the exact solution of
the short-time model, the bath is thermalized only initially,
while in the perturbative Markovian approximation, one assumes
that the bath is reset to thermal after each infinitesimal time
step. Nevertheless, the effect of the noise is expected to be
qualitatively similar. Since the short-time model offers an exact
solution, we will use it to compare the coherent vs noise effects
in the two-spin dynamics. We evaluate the
concurrence,\cite{Wootters1,Wootters2} which measures the
entanglement of the spin system and is monotonically related to
the entanglement of formation.\cite{Bennett,Vedral} For a mixed
state of two qubits, $\rho_S$, we first define the spin-flipped
state, $ \tilde \rho _S = \sigma^1_y \sigma^2_y \, \rho^*_S \,
\sigma^1_y \sigma^2_y $, and then the Hermitian operator
$R=\sqrt{\sqrt{\rho_S}\tilde\rho_S\sqrt{\rho_S}}$, with
eigenvalues $\lambda_{i=1,2,3,4}$. The concurrence is then
given\cite{Wootters2} by
\begin{equation}\label{eq:S3:concurence}
C\left( {\rho _S } \right) = \max \left\{ {0,2\mathop {\max
}\limits_i \lambda _i  - \sum\limits_{j = 1}^4 {\lambda _j } }
\right\}.
\end{equation}

In Figure~\ref{fig2}, we plot the concurrence as a function of
time and the spin-spin separation, for the (initially unentangled)
state $\left|\uparrow\uparrow\right\rangle$, and $n=1$. One
observes decaying periodic oscillations of entanglement. We should
point out that the measure of entanglement we use here --- the
concurrence --- provides the estimate of how much entanglement can
be constructed provided the worst possible scenario for
decomposing the density matrix is realized; see Refs.\
\onlinecite{Wootters1}, \onlinecite{Wootters2} for details and
definitions. Therefore, one expects the entanglement that one can
make use of in quantum computing to be no smaller than the one
presented in Figure~\ref{fig2}. In the next section, additional
quantities are considered, namely, the density matrix elements,
which characterize the degree of coherence in the system's
dynamics.

\section{Onset of the interaction and dynamics of the density
matrix} \label{Section4}

Let us now investigate in greater detail the onset of the induced
interaction, the time-dependence of which is given by $F(t)$. In
Figure~\ref{fig1} we have shown the magnitude of $F(t)$, as a
function of time for various qubit-qubit separations and $n=1$.
The correction is initially nonmonotonic, but decreases for larger
times as mentioned above. The behavior for other non-Ohmic regimes
is initially more complicated, however the large time behavior is
similar.

\begin{figure}
\includegraphics[width=8cm]{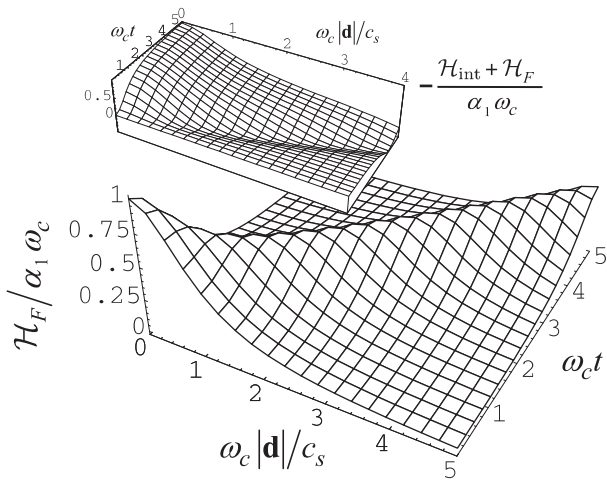}
\caption{The magnitude of the time-dependent Hamiltonian
corresponding to the initial correction as a function of time and
distance.  The Ohmic ($n=1$) case is shown. The inset demonstrates
the onset of the cross-qubit interaction on the same time
scale.}\label{fig4ad}
\end{figure}

It may be instructive to consider the time dependent correction
$H_F(t)$ to the interaction Hamiltonian during the initial
evolution, corresponding to $F(t)$. Since $F(t)$ commutes with
itself at different times, as well as with $H_\ind{int}$, it
generates unitary evolution according to $\exp[-i\int_0^t dt'
H_F(t')]$, with $H_F(t)=d[tF(t)]/dt$, therefore
\begin{eqnarray}\label{eq:H_F}
H_F(t)&=&\sigma _x^1 \sigma _x^2 \alpha_n\Gamma(n)
\\\nonumber
&\times&[%
u(\omega_c|\mathbf{d}|/c_s - \omega_c t)%
+u(\omega_c|\mathbf{d}|/c_s + \omega_c t)%
],
\end{eqnarray}
where $u(\xi)=\cos[n\arctan(\xi)]/[1+\xi^2]^{n/2}$. The above
expression is a superposition of two waves propagating in opposite
directions. In the Ohmic case, $n=1$, the shape of the wave is
simply $u(\xi)=1/(1+\xi^2)$. In Figure~\ref{fig4ad}, we present
the amplitude of $H_F(t)$, defined via $H_F(t)={\cal H}_F\sigma
_x^1 \sigma _x^2$, as well as the sum of $H_\ind{int}$ and
$H_F(t)$, for $n=1$. One can observe that the ``onset wave'' of
considerable amplitude and of shape $u(\xi)$ propagates once between
the qubits, ``switching on'' the interaction. It does not affect
the qubits once the interaction has set in.

To understand the dynamics of the qubit system and its
entanglement, let us again begin with the analysis of the coherent
part in Eq.(\ref{eq:S3:AdiabaticSolution}). After the interaction,
$H_\ind{int}$, has set in, it will split the system energies into
two degenerate pairs $E_0 = E_1=-{\cal H}_\ind{int}$ and $E_2 =
E_3={\cal H}_\ind{int}$. The wave function is then
$|\psi(t)\rangle=\exp(-iH_\ind{int}t)|\psi(0)\rangle$. For the
initial ``up-up'' state,
$|\psi(0)\rangle=\left|\uparrow\uparrow\right\rangle$, it develops
as $\left|\psi(t)\right\rangle=
\left|\uparrow\uparrow\right\rangle \cos {\cal H}_\ind{int}t +
\left|\downarrow\downarrow\right\rangle i\sin {\cal
H}_\ind{int}t$, where $H_\ind{int} = {\cal
H}_\ind{int}\sigma^1_m\sigma^2_m$. One can easily notice that at
times $t_E=\pi/4{\cal H}_\ind{int},3\pi/4{\cal
H}_\ind{int},\ldots$, maximally entangled (Bell) states are
obtained, while at times $t_0=0,\pi/2{\cal H}_\ind{int},\pi/{\cal
H}_\ind{int},\ldots$, the entanglement vanishes; these special
times can also be seen in Figure~\ref{fig2}. The coherent dynamics
obtained with the Markovian assumption is the same.

However, the coherent dynamics just described is only approximate,
because the bath also induces decoherence that enters via
Eq.(\ref{eq:S3:DecoherenceFunction-ReL}). The result for the
entanglement is that the decaying envelope function is
superimposed on the coherent ocsillations described above. The
magnitudes of the first and subsequent peaks of the concurrence
are determined only by this function. As temperature increases,
the envelope decays faster resulting in lower values of the
concurrence; see the inset in Figure~\ref{fig2}. Although in the
Markovian approach, presented in Section~\ref{Section2}, the noise
is quantitatively different, one expects qualitatively similar
results for the dynamics of entanglement.

Note also that nonmonotonic behavior of the entanglement is
possible only provided that the initial state is a superposition
of the eigenvectors of the induced interaction with more than one
eigenvalue (for pure initial states); see
Eqs.(\ref{eq:S3:AdiabaticSolution})-(\ref{eq:S3:DecoherenceFunction-ImL}).
For example, taking the initial state
$(\left|\downarrow\uparrow\right\rangle +
\left|\uparrow\downarrow\right\rangle)/\sqrt{2}$, in our case
would only lead to the destruction of entanglement, i.e., to a
monotonically decreasing concurrence, similar to the results of
Refs.\ \onlinecite{Eberly1}, \onlinecite{Eberly2}.

For the model that allows the exact solution, i.e., for $H_S=0$,
one can notice that there is no relaxation by energy transfer
between the system and bath. The exponentials in
Eqs.(\ref{eq:S3:AdiabaticSolution}), with
(\ref{eq:S3:DecoherenceFunction-ReL}), suppress only the
off-diagonal matrix elements, i.e., those with
$\lambda\neq\lambda'$. It happens, however, that at large times
the $\mathbf{d}$-dependence is not important in
Eq.(\ref{eq:S3:DecoherenceFunction-ReL}), and
$\textrm{Re}\frak{L}_{\lambda\lambda'}(t\rightarrow\infty)$
vanishes for certain values of $\lambda\neq\lambda'$. In the basis
of the qubit-bath interaction, $\sigma_x^1\sigma_x^2$, the
limiting $t\to \infty$ density matrix for our initial state
($\left|\uparrow\uparrow\right\rangle$) is
$\frac{1}{4}+\frac{1}{4}\left|+-\right\rangle\left\langle-+\right|
+\frac{1}{4}\left|-+\right\rangle\left\langle+-\right|$, which
takes the form
\begin{equation}\label{eq:roLimit}
\rho(t\rightarrow\infty)\rightarrow\frac{1} {8}\left( {
{\begin{array}{*{20}c}
   3 & 0 & 0 & { - 1}  \\
   0 & 1 & 1 & 0  \\
   0 & 1 & 1 & 0  \\
   { - 1} & 0 & 0 & 3  \\
 \end{array} }} \right)
\end{equation}
in the basis of states $\left|\uparrow\uparrow\right\rangle$,
$\left|\uparrow\downarrow\right\rangle$,
$\left|\downarrow\uparrow\right\rangle$, and
$\left|\downarrow\downarrow\right\rangle$. The significance of
such results, see also Ref.~\onlinecite{Braun}, is that in the
model with $H_S=0$ and nonrethermalizing bath not all the
off-diagonal matrix elements are suppressed by decoherence, even
though the concurrence of this mixed state is zero.

\begin{figure}
\includegraphics[width=5cm]{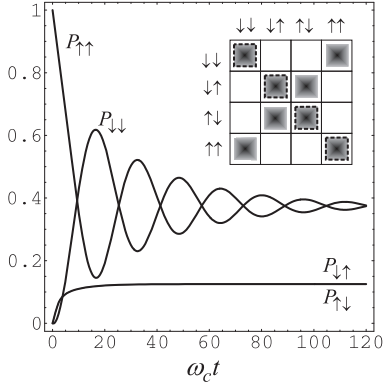}
\caption{Dynamics of the occupation probabilities for the states
$\left|\uparrow\uparrow\right\rangle$,
$\left|\uparrow\downarrow\right\rangle$,
$\left|\downarrow\uparrow\right\rangle$, and
$\left|\downarrow\downarrow\right\rangle$. The parameters are the
same as in Figure~\ref{fig2}. The inset shows the structure of the
reduced density matrix (the nonshaded entries are all
zero).}\label{fig5ad}
\end{figure}

\begin{figure}
\includegraphics[width=6cm]{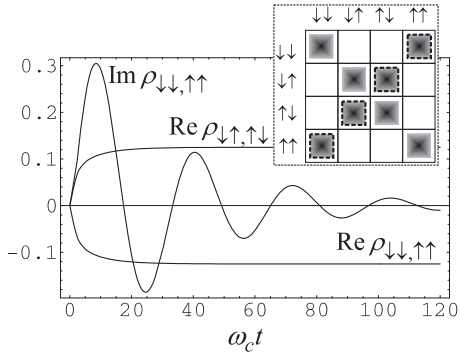}
\caption{Dynamics of the inverse-diagonal matrix elements for the same
system as in Figure~\ref{fig5ad}. Note that ${\rm Im}\,\rho_{\downarrow\uparrow,\uparrow\downarrow}=0$.}\label{fig6ad}
\end{figure}

The probabilities for the spins to occupy the states
$\left|\uparrow\uparrow\right\rangle$,
$\left|\uparrow\downarrow\right\rangle$,
$\left|\downarrow\uparrow\right\rangle$, and
$\left|\downarrow\downarrow\right\rangle$ are presented in
Figure~\ref{fig5ad}. For the initial state
$\left|\uparrow\uparrow\right\rangle$, only the diagonal and
inverse-diagonal matrix elements are affected, and the system
oscillates between the two states
$\left|\uparrow\uparrow\right\rangle$ and
$\left|\downarrow\downarrow\right\rangle$, as mentioned earlier in
the description of the coherent dynamics, while decoherence
dampens these oscillations. In addition, decoherence actually
raises the other two diagonal elements to a certain level, see
Eq.(\ref{eq:roLimit}). The dynamics of the inverse-diagonal
density matrix elements is shown in Figure~\ref{fig6ad}.

\section{Ohmic and Super-Ohmic Bath Models in General Dimension} \label{Section5}
Let us generalize the results of the previous sections obtained
primarily for the Ohmic bath model and 1D geometry. In
Section~\ref{Section2}, we considered the 1D case with Ohmic
dissipation with the Markovian approach. In the general case, let
us consider the Markovain model and, again, assume that
$\Delta/\omega_c$ is small. We will also assume that the absolute
square of the $m$th component of the spin-boson coupling, when
multiplied by the density of states, can be modelled by
$\alpha_{n_m}^m{}\omega^{n_m}\exp(-\omega/\omega_c)$; see
Eq.(\ref{eq:S2:DOS}). The integration in Eq.(\ref{eq:S2:he-c}) can
then be carried out in closed form for any $n_m=1,2,\ldots\,\,$.
The induced interaction (\ref{eq:S2:H-int-1D}) is thus generalized
to
\begin{eqnarray}\label{eq:S4:H-int}
H_\textrm{int}=&-&\!\!\!\sum\limits_{m =
x,y,z}{\alpha_{n_m}^m\omega_c^{n_m }\sigma_m^1\sigma _m^2}
\\ \nonumber
&\times&\frac{{2\Gamma (n_m
)\textrm{Re}\left({1+i\omega_c\left|{\mathbf{d}}\right|/c_s}\right)^{n_m}}}
{{\left[{1+\left({\omega_c\left|{\mathbf{d}}\right|/c_s}\right)^2}\right]^{n_m}}}.
\end{eqnarray}
With the appropriate choice of parameters, the result for the
induced interaction, but not for the noise, coincides with the
expression for $H_\textrm{int}$ obtained within the short-time
model. From Eq.(\ref{eq:S4:H-int}) one can infer that the
effective interaction has the large-distance asymptotic behavior
$\left|{\mathbf{d}}\right|^{-n_m}$, for even $n_m$, and
$\left|{\mathbf{d}}\right|^{-n_m-1}$, for odd $n_m$.
\begin{figure}
\includegraphics[width=8cm]{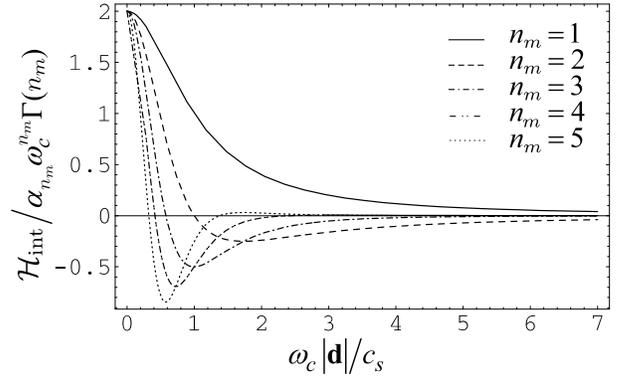}
\caption{The magnitude of the induced exchange interaction
Hamiltonian for the Ohmic $n_m=1$ and super-Ohmic $n_m=2,3,4,5$
baths.}\label{fig3}
\end{figure}
\begin{figure}
\includegraphics[width=8cm]{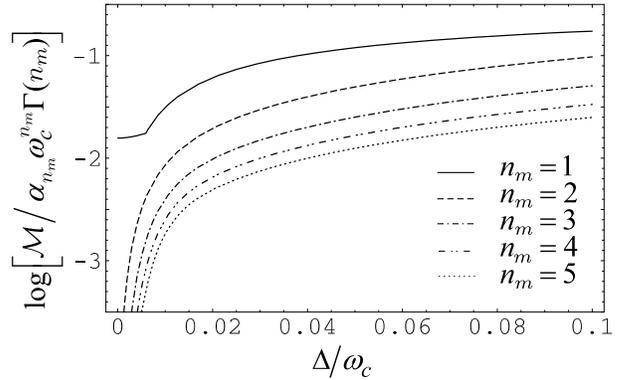}
\caption{The magnitude of the noise for the Ohmic $n_m=1$ and
super-Ohmic $n_m=2,3,4,5$ baths, for
$2k_\textrm{B}T/\omega_c=0.01$.}\label{fig4}
\end{figure}

Let us now consider the noise terms. The amplitudes
(\ref{eq:S2:eta-c}, \ref{eq:S2:he-s}), entering the decoherence
superoperator, depend on $n_m$ only via the prefactor
$(\Delta/\omega_c)^{n_m}$. The remaining amplitude
(\ref{eq:S2:eta-s}) has to be calculated numerically. The
spin-spin separation up to which the ``coherent'' effects of the
induced interaction can be observed in the time-dependent
dynamics, is defined by comparing the magnitudes of the
interaction and noise amplitudes. It transpires that the
comparison of the amplitude in Eq.(\ref{eq:S2:H-int-1D}) vs\
Eq.(\ref{eq:S2:M-explicit-1D}) for $n_m=1$, also suffices for an
approximate estimate for the super-Ohmic case, as long as
$\eta^m_s(\mathbf{d})$ does not dominate in ${\hat M}$. In
Figure~\ref{fig3}, we plot the amplitudes of
Eq.(\ref{eq:S4:H-int}), i.e., ${\cal H}_\textrm{int}\equiv
Tr(H_\textrm{int}\sigma^1_m\sigma^2_m)/4$, for different values of
$n_m$. Figure~\ref{fig4} shows the magnitude of the noise
super-operator. Figures~6 and 7 suggest that the noise amplitudes
can be made sufficiently small with respect to the induced
interaction, at small values of $\Delta/\omega_c$, for any $n_m$.
We note that for quantum computing applications one usually
assumes the regime $k_BT \ll \Delta \ll \omega_c$. The temperature
dependence in Fig.~7 becomes insignificant for $k_BT/\Delta \ll
1$, which is approximately to the right of $\Delta/\omega_c \sim
0.01$. To the left of $\Delta/\omega_c \sim 0.01$, by reducing the
temperature one can further reduce the values of the noise
amplitudes even for small $\Delta$.

In higher dimensions the structure of
$g_{{\mathbf{k}},\mathbf{\xi}}^m$ in the $\textbf{k}$-space
becomes important. Provided $\omega _{{\mathbf{k}},\mathbf{\xi}}$
is nearly isotropic, the integrals entering
Eqs.(\ref{eq:S2:he-c})-(\ref{eq:S2:eta-s}) will include (in 3D) a
factor
$\int_0^{2\pi}{d\varphi}\int_0^\pi{d\theta\sin\theta}\left|{
g_{k\theta\varphi,\mathbf{\xi}}^m}\right|^2\cos\left({k\left|{\mathbf{d}}
\right|\cos\theta}\right)$, which can be written as
$[f^m_1(\omega,k|\mathbf{d}|)-f^m_2(\omega,
k|\mathbf{d}|)\partial/\partial|\mathbf{d}|]\cos (k|\mathbf{d}|)$,
e.g., Eqs.(\ref{eq:B:I-xy}, \ref{eq:B:I-z}) in
Appendix~\ref{SecB}. When the dependence of $f^m_1,f^m_2$ on
$k|\mathbf{d}|$ is negligible, the interaction is simply
$H_\textrm{int}\rightarrow H_\textrm{int}|_{\{n_m\}\rightarrow
a}-(\partial/\partial|\mathbf{d}|)H_\textrm{int}|_{\{n_m\}\rightarrow
b}$, where $a$ and $b$ are sets of three integers representing the
$\omega$-dependence of $f_1^m(\omega,k|\mathbf{d}|)$ and
$f_2^m(\omega,k|\mathbf{d}|)$. Otherwise, a more complicated
dependence on $|\mathbf{d}|$ is expected. The noise superoperator
can be treated similarly. As a result one can see that the form of
the interaction depends more on the structure of
$g_{{\mathbf{k}},\mathbf{\xi}}^m$ than on the dimensionality via
the phonon density of states. However, reduced dimensionality in
the $\mathbf{k}$-space might allow for better control over the
magnitude of the interaction by external potentials that modify
the spin-orbit coupling.

\section{Phonon Induced Coherent Spin-Spin Interaction vs Noise for
P-donor electrons in $\textrm{Si}$ and $\textrm{Ge}$} \label{Section6}
As a specific example, let us consider a model of two localized
impurity-electron spins of phosphorus donors in a Si-Ge type
semiconductor, coupled to acoustic phonon modes by spin-orbit
interaction. In what follows, we first obtain the coupling
constants $g^m_{\mathbf{k}, \xi}$, entering Eq.(\ref{eq:S1:X_jm}),
which define the interaction and noise amplitudes. A brief
discussion is then offered on the possibility of having an Ohmic
bath model realized in 1D channels. The rest of the present
section is devoted to calculations of the induced interaction and
noise in 3D bulk material. A comparison with the dipole-dipole
spin interaction is given.

Averaging the spin orbit Hamiltonian over the localized electron's
wave function, one obtains the spin-orbit coupling in the presence
of magnetic field $\mathbf{H}=(H_x,H_y,H_z)$ in the form
\begin{equation}\label{eq:S5:H-SO}
H_\textrm{SO}=\mu_{\textrm{B}} \! \! \sum\limits_{m,l\,=\,x,y,z} \! \! {\sigma_m g_{ml}H_l},
\end{equation}
where $\mu_\textrm{B}$ is the Bohr magneton. Here the tensor
$g_{ml}$ is sensitive to lattice deformations. It was
shown\cite{Roth} that for the donor state which has tetrahedral
symmetry, the Hamiltonian (\ref{eq:S5:H-SO}) yields the
spin-deformation interaction of the form
\begin{eqnarray}\nonumber
H &=& {\rm A}\mu _{\textrm{B}} \left[ { \bar \varepsilon _{xx} \sigma _x H_x
+ \bar \varepsilon _{yy} \sigma _y H_y  +  \bar
\varepsilon _{zz} \sigma _z H_z  + {\bar\Delta }
(\mathbf{\sigma}\cdot\mathbf{H}})/3\right]
\\ \label{eq:S5:H-SD-Hxyz}
&+&{\rm B}\mu _{\textrm{B}} \left[ {\bar \varepsilon _{xy} \left({ \sigma_x
H_y+\sigma_y H_x}\right)+{\text{c}}{\text{.p}}{\text{.}}}\right].
\end{eqnarray}
Here c.p.\ denotes cyclic permutations and $\bar\Delta$ is the
effective dilatation. The tensor $\bar\varepsilon_{ij}$ already
includes averaging of the strain with the gradient of the
potential over the donor ground state wave function.

As before, let us assume that the separation \textbf{d}, as well
as the magnetic field, are directed along the $z$-axis, for an
illustrative calculation. Then the spin-deformation interaction
Hamiltonian simplifies to
\begin{equation}\label{eq:S5:H-SD-Hz}
H = {\rm A}\mu _{\textrm{B}}  \bar \varepsilon _{zz} \sigma _z H_z  + {\rm
B}\mu _{\textrm{B}} \left( {\bar \varepsilon _{yz} \sigma _y H_z  + \bar
\varepsilon _{zx} \sigma _x H_z } \right).
\end{equation}
In terms of the quantized phonon field, we have\cite{Mahan,MKGB}
\begin{equation}\label{eq:S5:StrainTensor}
\bar \varepsilon _{ij}  = \sum\limits_{{\mathbf{k}},{\mathbf{\xi
}}} {f({\mathbf{k}})\sqrt {\frac{\hbar } {{8\rho V\omega
_{{\mathbf{k}},{\mathbf{\xi }}} }}} \left( {\xi _{{\mathbf{k}},i}
k_j  + \xi _{{\mathbf{k}},j} k_i } \right)\left(
{a_{{\mathbf{k}},{\mathbf{\xi }}}^\dag   +
a_{{\mathbf{k}},{\mathbf{\xi }}} } \right)},
\end{equation}
where in the spherical donor ground state
approximation\cite{Hasegawa,MKGB}
\begin{equation}\label{eq:S5:f(k)}
f({\mathbf{k}}) = \frac{1} {{\left( {1 + a_\textrm{B}^2 k^2 }
\right)^2}}.
\end{equation}
Here $a_\textrm{B}$ is {\it half\/} the effective Bohr radius of
the donor ground state wave function. In an actual Si or Ge
crystal, donor states are more complicated and include corrections
due to the symmetry of the crystal matrix including the fast
Bloch-function oscillations. However, the wave function of the
donor electrons in our case is spread over several atomic
dimensions (see below). Therefore, it suffices to consider
``envelope'' quantities. Thus, the spin-phonon Hamiltonian
(\ref{eq:S1:H_SB}, \ref{eq:S1:X_jm}) coupling constants will be
taken in the form
\begin{equation}\label{eq:S5:gk-SD}
g_{{\mathbf{k}},{\mathbf{\xi }}}^m  = \frac{D_m}{{\left( {1 +
a_\textrm{B}^2 k^2 } \right)^2}}\sqrt {\frac{\hbar } {{8\rho
V\omega _{{\mathbf{k}},{\mathbf{\xi }}} }}} \left( {\xi
_{{\mathbf{k}},z} k_m  + \xi _{{\mathbf{k}},m} k_z } \right),
\end{equation}
where $D_x=D_y={\rm B}\mu_\textrm{B} H_z$ and $D_z={\rm
A}\mu_\textrm{B} H_z$.

\begin{figure}
\includegraphics[width=8cm]{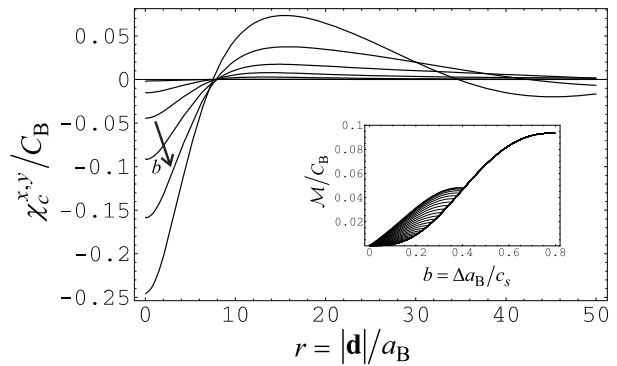}
\caption{The magnitude of the induced spin-spin interaction for a
3D Si-Ge type structure: The dominant interaction amplitude, which
is the same for the $x$ and $y$ spin components, is shown. The
arrow indicates increasing $b$ values for the curves shown, with
$b=0.03$, $0.09$, $0.15$, $0.21$, $0.27$, $0.33$. The inset
estimates the level of the noise (for $2k_BTa_B/c_s=0.01$): The
bottom curve is $\eta_c^{m}(0)$, with $m=x,y$. For $b \lesssim
0.4$, the amplitude $\eta_s^{m}(\mathbf{d})$ can be comparable,
and its values, calculated numerically for $0\leq r\leq 50$, are
shown as long as they exceed $\eta_c^{m}(0)$, with the top curve
corresponding to the maximum value, at $r=0$.}\label{fig5}
\end{figure}
Let us first consider a 1D channel geometry along the $z$
direction. This will give an example of an Ohmic bath model
discussed towards the end of Section~\ref{Section2}. In 1D channel
the boundaries\cite{PBS} can approximately quantize the spectrum
of phonons along $x$ and $y$, depleting the density of states
except at certain resonant values. Therefore, the low-frequency
effects, including the induced coupling and quantum noise, will
become effectively onedimensional, especially if the effective gap
due to the confinement is of the order of $\omega_c$. As mentioned
earlier, this frequency cutoff comes from Eq.(\ref{eq:S5:f(k)}),
namely, it is due to the bound electron wave function
localization. A channel of width comparable to $\sim a_\textrm{B}$
will be required. This, however, may be difficult to achieve in
bulk Si or Ge with the present-day technology. Other systems may
offer more immediately available 1D geometries for testing similar
theories, for instance, carbon nanotubes, chains of ionized atoms
suspended in ion traps,\cite{Porras,Marquet,Leibfried} etc. In our
case, the longitudinal acoustic (LA, $||$) phonons of the
one-dimensional spectrum will account for the
$g_{\mathbf{k},\parallel}^z$ component of the coupling, whereas
the transverse acoustic (TA, $\perp$) phonons will affect only the
$x$ and $y$ spin projections.

One can show that the contributions of the crossproducts of
coupling constants,
$g_{\mathbf{k},\mathbf{\xi}}^m(g_{\mathbf{k},\mathbf{\xi}}^{m'})^*$
with $m\neq m'$, to quantities of interest vanish. The diagonal
combinations are
\begin{eqnarray}\label{eq:S5:gk-1D}
|g_{k_z }^z|^2 &=& \frac{\textrm{A}^2 \mu_\textrm{B}^2 H_z^2}
{{4\rho V\omega _{k_z ,\parallel } }}\frac{{k_z^2 }} {{\left( {1 +
a_\textrm{B}^2 k_z^2 } \right)^4 }},
\\ \nonumber
|g_{k_z }^x |^2  &=& |g_{k_z }^y |^2  = \frac{\textrm{B}^2
\mu_\textrm{B}^2 H_z^2} {{4\rho V\omega _{k_z , \bot }
}}\frac{{k_z^2 }} {{\left( {1 + a_B^2 k_z^2 } \right)^4 }}.
\end{eqnarray}
With our usual assumption for the low-frequency dispersion
relations $\omega _{k_z ,\parallel}\approx c_\parallel k_z$ and
$\omega _{k_z ,\perp}\approx c_\perp k_z$, the expressions
(\ref{eq:S5:gk-1D}) lead to the Ohmic bath model discussed at the
end of Section~\ref{Section2}. The shape of the frequency cutoff
resulting from Eq.(\ref{eq:S5:gk-1D}) is not exponential. However,
to estimate the magnitude of the interaction and noise one can
utilize the results obtained earlier for the 1D Ohmic bath model
with exponential cutoff. The coupling constants in
Eqs.(\ref{eq:S2:H-int-1D}) and (\ref{eq:S2:M-explicit-1D}) should,
then, be taken as
\begin{eqnarray}\label{eq:S5:alfas-1D}
\alpha_1^z &=& \frac{\textrm{A}^2 \mu_\textrm{B}^2 H_z^2}{8\pi
\rho S c^3_\parallel},
\\ \nonumber
\alpha_1^x  &=& \alpha_1^y  = \frac{\textrm{B}^2 \mu_\textrm{B}^2
H_z^2}{8\pi \rho S c^3_\bot},
\end{eqnarray}
where $S$ is a cross section of the channel, and the cutoff is
$\omega_c \rightarrow c_{\parallel}/a_\textrm{B}$ for the $z$
component, and $\omega_c \rightarrow c_{\perp}/a_\textrm{B}$ for
the $x$ and $y$ components. Considering Si as an example, we
arrive at an approximately adiabatic Hamiltonian ($\alpha_1^z \gg
\alpha_1^{x,y}$) with Ohmic-type coupling. The dynamics of the
concurrence, then, is qualitatively similar to the one shown on
Fig.~2, with the peak entanglement $\sim 0.4$ as well. The
coupling constant $\alpha_1^z$, however, is significantly smaller
due to the weakness of the spin-orbit coupling of P-impurity
electrons in Si, which results in low magnitude of the induced
interaction (and the noise due to the same environment), and
slightly modifies the shape of the concurrence.

In the 3D geometry, let us consider for simplicity only the LA
phonon branch, ${\mathbf{\xi }} \to {\mathbf{k}}/\left|
{\mathbf{k}} \right|$, and assume an isotropic dispersion
$\omega_{\mathbf{k},\mathbf{\xi}}=c_s\left|\mathbf{k}\right|$. The
expression for the coupling constants is then
\begin{equation}\label{eq:S5:gk-3D}
g_{{\mathbf{k}},{\mathbf{\xi }}}^m  = D_m \frac{{k_z k_m }}
{{\left( {1 + a_\textrm{B}^2 k^2 } \right)^2 }}\sqrt {\frac{\hbar
} {{2\rho Vc_s k^3 }}}.
\end{equation}
The cross terms, with $m \neq n$, of the correlation functions
$Tr_B \left[ X_m^j X_n^i (t)\rho_B \right]$, see
Eq.(\ref{eq:A:C-explicit}) in Appendix~\ref{SecA}, depend on the
combination $g_{\mathbf{k}, \mathbf{\xi}}^m(g_{\mathbf{k},
\mathbf{\xi}}^n)^*$, which is always an odd function of  one of
the projections of the wave vector. The nondiagonal terms thus
vanish, as mentioned in Appendix~\ref{SecA}.

Integrating Eqs.(\ref{eq:S2:he-c}) and (\ref{eq:S2:eta-c}) with
Eq.(\ref{eq:S5:gk-3D}), see Appendix~\ref{SecB}, one can
demonstrate that decoherence is dominated by the individual noise
terms for each spin, with the typical amplitude
\begin{equation}\label{eq:S5:eta-c-xy}
\eta _c^{x,y} (0) = C_\textrm{B} \frac{{2\pi ^2 }}
{{15}}\frac{b^3}{\left(
{1+b^2}\right)^4}\coth\frac{\Delta}{2k_\textrm{B}T},
\end{equation}
where $b = \Delta a_B /c_s$ and $C_\textrm{B}={\rm B}^2\mu_B^2
H_z^2/\left({16\pi^3\rho \hbar a_B^3 c_s^2}\right)$. The
interaction amplitude $\chi _c^m ({\mathbf{d}})$ and, therefore,
the induced spin-spin interaction, has inverse-square power-law
asymptotic form for the $x$ and $y$ spin components, with a
superimposed oscillation, and inverse-fifth-power-law decay for
the $z$ spin components
\begin{eqnarray}\label{eq:S5:H-int-asymptotic}
H_\textrm{int}&=&\sum\nolimits_m {2\chi _c^m ({\mathbf{d}})\sigma
_m^1 \sigma _m^2 }
\\ \nonumber
\xrightarrow{{r\gg1}} &-& 4\pi ^2 C_\textrm{B} \frac{{2b}}
{{\left( {1 + b^2 } \right)^4 }}\frac{{\sin br}} {{r^2 }}\left(
{\sigma _x^1 \sigma _x^2  + \sigma _y^1 \sigma _y^2 } \right)
\\ \nonumber
&+& 384\pi ^2 C_\textrm{A} \frac{2} {{r^5 }}\sigma _z^1 \sigma
_z^2.
\end{eqnarray}
Here $r=|\mathbf{d}|/a_\textrm{B}$ and
$C_\textrm{A}=\textrm{A}^2C_\textrm{B}/\textrm{B}^2$. At small
distances the interaction is regular and the amplitudes converge
to constant values, see Figure~\ref{fig5}. The complete
expressions for $\chi_c^m(\mathbf{d})$ and $\eta_c^m(\mathbf{d})$
are given in Appendix~\ref{SecB}.

In Figure~\ref{fig5}, we plot the amplitudes of the induced
spin-spin interaction (\ref{eq:S5:H-int-asymptotic},
\ref{eq:B:he-xy-c}, \ref{eq:B:he-z-c}) and noise for different
values of the spin-spin separation and $b$, for electron impurity
spins in 3D Si-Ge type structures. The value of $b$ can be
controlled via the applied magnetic field, $b=\mu_\textrm{B}H_z
g^* a_\textrm{B}/c_s$. The
temperature dependence of the noise is insignificant provided $2k_BT/\Delta\ll 1$.

A typical value\cite{Hasegawa,Roth,MKGB} of the effective Bohr
radius in Si for the P-donor-electron ground state wave function
is $2a_\textrm{B}=2.0\,$nm. The crystal lattice density is
$\rho=2.3\times10^3\,$kg/m$^3$, and the g-factor $g^*=1.98$. For
an order-of-magnitude estimate, we take a typical value of the
phonon group velocity, $c_s=0.93 \times 10^4\,$m/s. The spin-orbit
coupling constants in Si are\cite{Hasegawa,Roth,MKGB}
$\textrm{A}^2\approx 10^2$ and $\textrm{B}^2\approx10^{-1}$. The
resulting interaction constants in Eqs.(\ref{eq:S5:eta-c-xy},
\ref{eq:S5:H-int-asymptotic}) are $C_\textrm{B}=7.8\,$s$^{-1}$ and
$C_\textrm{A}=10^3C_\textrm{B}$. In the Ge lattice, the spin-orbit
coupling is dominated by the non-diagonal
terms,\cite{Hasegawa,Roth,MKGB} $\textrm{A}^2\approx 0$ and
$\textrm{B}^2\approx10^6$. The other parameters are
$2a_\textrm{B}=4.0\,$nm, $\rho=5.3\times10^3\,$kg/m$^3$,
$c_s=5.37\times10^3\,$m/s, and $g^* = 1.56$. This results in a
much stronger transverse component interaction,
$C_\textrm{B}=1.3\times{}10^7\,$s$^{-1}$ and
$C_\textrm{A}\approx0$. In both cases the magnetic field was taken
$H_z=3\times 10^4\,$G. In the above estimations, one could use
various experimentally suggested values for the parameters, such
as, for instance, $a_B$. This will not affect the results
significantly.

\begin{figure}
\includegraphics[width=8cm]{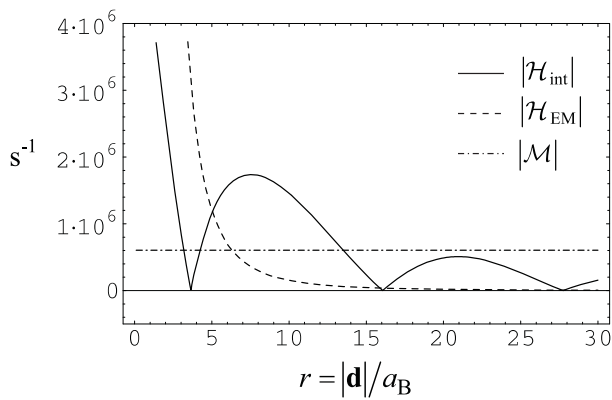}
\caption{The magnitudes, measured in units of s$^{-1}$, of the
induced spin-spin interaction, the EM coupling strength, and the
level of noise for P-impurity electron spins in Ge. Here $H_z=3
\times 10^4\,$G, and low temperature, $2k_\textrm{B}T/\Delta \ll
1$, was assumed.}\label{fig6}
\end{figure}
As mentioned is Section~\ref{Section2}, the obtained interaction
(\ref{eq:S5:H-int-asymptotic}) is always accompanied by noise
coming from the same source, as well as possibly by other, direct
interactions of the spins. When the electron wave functions
overlap is negligible, the dominant direct interaction will be the
dipole-dipole one
\begin{equation}\label{eq:S5:EM-spin-spin-interaction}
H_\textrm{EM}(\mathbf{d})=\frac{\mu_0 \mu _\textrm{B}^2}{4\pi}
\frac{\sigma _x^1 \sigma _x^2  + \sigma _y^1 \sigma _y^2  -
2\sigma _z^1 \sigma _z^2}{|\mathbf{d}|^3}.
\end{equation}
The comparison of the two interactions and noise is shown in
Figure~\ref{fig6}. We plot the magnitude of the effective induced
interaction (\ref{eq:S5:H-int-asymptotic}, \ref{eq:B:he-xy-c},
\ref{eq:B:he-z-c}), the electromagnetic interaction
(\ref{eq:S5:EM-spin-spin-interaction}), measured by ${\cal
H}_\textrm{EM}\equiv{\mu_0 \mu _\textrm{B}^2}/{4\pi
|\mathbf{d}|^3}$, and a measure of the level of noise, for P-donor
electron spins in Ge. It transpires that the induced interaction
can be considerable as compared to the electromagnetic spin-spin
coupling. However the overall coherence-to-noise ratio is quite
poor for Ge. In Si, the level of noise is lower as compared to the
induced interaction. However, the overall amplitude of the induced
terms compares less favorably with the electromagnetic coupling.

In conclusion, we have studied the induced indirect exchange
interaction due to a bosonic bath which also introduces quantum
noise. We demonstrated that it can create substantial two-spin
entanglement. For an appropriate choice of the system parameters,
specifically, the spin-spin separation, for low enough
temperatures this entanglement can be maintained and the system
can evolve approximately coherently for many cycles of its
internal dynamics. For larger times, the quantum noise effects
will eventually dominate and the entanglement will be erased.

Estimates for P-impurity electron spins in Si and Ge structures
have demonstrated that the induced interaction can be comparable
to the dipole-dipole spin interaction. One can also infer that
this phonon-mediated interaction in the bulk (3D) Ge is not very
effective to be used for quantum computing purposes, i.e., to
entangle qubits. This is due to poor coherence to noise ratio. Therefore,
the use of Si may be favored, despite the fact that it has weaker
spin-phonon coupling. Indeed, the noise amplitudes for Si are
significantly smaller then the induced exchange
interaction, where the latter is
dominated by the adiabatic term. The situation is
expected to be further improved for systems with reduced dimensionality
for phonon propagation.

{\acknowledgements The authors acknowledge useful discussions with
and helpful suggestions by J.~Eberly, L.~Fedichkin, D.~Mozyrsky,
and I.~Shlimak. This research was supported by the NSF under Grant
No. DMR-0121146.}

\appendix
\renewcommand{\theequation}{\thesection .\arabic{equation}}

\section{Derivation Steps for the Induced Interaction and Noise}\label{SecA}
An important assumption required for the validity of the Markovian
approach concerns the time scale of the decay of the bath
correlations introduced in Eqs.(\ref{eq:A:CviaTrXX}) and
(\ref{eq:A:C-explicit}) below. By constantly resetting the bath to
thermal, one implies that this time scale is significantly shorter
than the dynamical system time scales of interest. The Markovian
treatment, considered here, is, therefore, valid at all but very
short times. The short-time dynamics, for the time scales down to
order $1/\omega_c$, requires a different
approach.\cite{Privman,Tolkunov2,Solenov} Note that one usually
assumes that $\omega_c \gg \Delta$.

Here we review some of the steps that lead from the Markovian
equation for the density matrix (\ref{eq:S2:MME}), to the
expressions for the induced interaction and quantum noise.
Substituting Eqs.(\ref{eq:S1:H_SB}) and (\ref{eq:S1:X_jm}) in
Eq.(\ref{eq:S2:MME}), one can represent the integrand, $Tr_B
[H_{SB} ,[H_{SB} (t' - t),\rho _B \rho _S (t)]]\rho _S (t)$, as a
summation over $i,j,m,n$, of the following expression:
\begin{eqnarray}\nonumber
 \hphantom{+} Tr_B \left( {X^j_m X^i_n (t' - t)\rho _B } \right)\sigma^j_m \sigma^i_n (t' - t)\rho_S (t)
\\ \label{eq:A:MME-inteagrand}
 - Tr_B \left( {X^j_m \rho _B X^i_n (t' - t)} \right)\sigma^j_m \rho_S (t)\sigma^i_n (t' - t)
\\ \nonumber
 - Tr_B \left( {X^i_n (t' - t)\rho _B X^j_m } \right)\sigma^i_n (t' - t)\rho_S (t)\sigma^j_m
\\ \nonumber
 + Tr_B \left( {\rho _B X^i_n (t' - t)X^j_m } \right)\rho_S (t)\sigma^i_n (t' - t)\sigma^j_m .
\end{eqnarray}
Here $\sigma_n^i (t)=e^{iH_St}\sigma_n^i{}e^{-iH_St}$. All the
terms in Eq.(\ref{eq:A:MME-inteagrand}) involve the correlation
functions
\begin{equation}\label{eq:A:CviaTrXX}
C_{mn} \left( {(1 - \delta _{ij} ){\mathbf{d}},t} \right) = Tr_B
X_m^j X_n^i (t)\rho _B,
\end{equation}
where $X_n^i (t)=e^{iH_Bt}X_n^i{}e^{-iH_Bt}$. The explicit
expression for these functions can then be obtained from
Eqs.(\ref{eq:S1:X_jm}) and (\ref{eq:A:CviaTrXX}),
\begin{eqnarray}\label{eq:A:C-explicit}
  C_{mn} ({\mathbf{d}},t) = \frac{V}
{{\left( {2\pi } \right)^3 }}\sum\limits_\mathbf{\xi} \!
{\int\limits_{ - \infty }^\infty \! \! {d{\mathbf{k}}} } \left|
{g_{{\mathbf{k}},\mathbf{\xi}}^m g_{{\mathbf{k}},\mathbf{\xi}}^n }
\right| \Big[ {i\sin \omega _{{\mathbf{k}},\mathbf{\xi}}
t}\phantom{\frac{a}{b}}
\\ \nonumber
+ \, {\coth \frac{{\omega _{{\mathbf{k}},\mathbf{\xi}} }}
{{2k_\textrm{B}T}}\cos \omega _{{\mathbf{k}},\mathbf{\xi}} t}
\Big]\cos \left( {{\mathbf{k}} \cdot {\mathbf{d}} + \phi
_{{\mathbf{k}},\mathbf{\xi}}^n - \phi
_{{\mathbf{k}},\mathbf{\xi}}^m } \right),
\end{eqnarray}
where $\phi_{\mathbf{k},\xi}^n$ is a possible phase of the
coupling constants $g_{\mathbf{k},\xi}^n$, which is not present in
most cases. The coupling constants are examined in detail in
Sections~\ref{Section5} and \ref{Section6}, and explicit model expressions are given.
Presently, we only comment that in many cases the resulting matrix
of the correlation functions $C_{mn} ({\mathbf{d}},t)$ is
diagonal, which simplifies calculations, as illustrated in Section~\ref{Section6}.

The summation of Eq.(\ref{eq:A:MME-inteagrand}) over $i,j,m,n$ is
further simplified by noting that $C_{mn}(\mathbf{d},t) =
C^{*}_{mn}(\mathbf{d},-t)$, and writing $\sigma_n^i(t)$ explicitly
as $\sigma_z^i$ for $n=z$, and $\sigma_n^i\cos\Delta t +
\frac{1}{2} [\sigma_z^i,\sigma_n^i]\sin\Delta t$ for $n\neq z$.
For a diagonal $C_{m,n}$, we then get the amplitude expressions
\begin{eqnarray}
\int\limits^0_{-\infty}dt'
\textrm{Im}C_{mm}(\mathbf{d},t')\cos(\Delta
t')=\chi_c^m(\mathbf{d}) ,
\\
\int\limits^0_{-\infty}dt'
\textrm{Im}C_{mm}(\mathbf{d},t')\sin(\Delta
t')=\chi_s^m(\mathbf{d}) ,
\\
\int\limits^0_{-\infty}dt'
\textrm{Re}C_{mm}(\mathbf{d},t')\cos(\Delta
t')=\eta_c^m(\mathbf{d}) ,
\\
\int\limits^0_{-\infty}dt'
\textrm{Re}C_{mm}(\mathbf{d},t')\sin(\Delta
t')=\eta_s^m(\mathbf{d}) .
\end{eqnarray}
This finally leads to Eqs.(\ref{eq:S2:H-eff}) and
(\ref{eq:S2:M-sums}).

\section{Interaction and Noise Amplitudes for
$\textrm{Si}$-$\textrm{Ge}$ type Spin-Orbit Coupling}\label{SecB}
In Eqs.(\ref{eq:S2:he-c}) and (\ref{eq:S2:eta-c}), with
Eq.(\ref{eq:S5:gk-3D}), there is a common angular part
\begin{equation}\label{eq:B:I-integral}
I_m(kd) = \int\limits_0^{2\pi}d\varphi \int\limits_0^{\pi}
\sin\theta d\theta \frac{k_z^2 k_m^2}{k^4} cos(kd \cos\theta),
\end{equation}
which gives
\begin{eqnarray}\label{eq:B:I-xy}
I_{x,y}(kd) &=& 4\pi \operatorname{Re} \frac{{12 - k^2d^2 }}
{{k^4d^4 }}e^{ikd}
\\ \nonumber
 &-& 4\pi \operatorname{Re} i\frac{{5k^2d^2  - 12}} {{k^5d^5
}}e^{ikd}
\end{eqnarray}
and
\begin{eqnarray}\label{eq:B:I-z}
I_{z}(kd) &=& 16\pi \operatorname{Re} \frac{{k^2d^2  - 6}}
{{k^4d^4 }}e^{ikd}
\\ \nonumber
&-& 4\pi \operatorname{Re} i\frac{{k^4d^4  - 12k^2d^2  + 24}}
{{k^5d^5 }}e^{ikd}.
\end{eqnarray}
Here $d\equiv |\mathbf{d}|$. The remaining integral in
Eq.(\ref{eq:S2:he-c}),
\begin{equation}\label{eq:B:he-xy-c}
\chi _c^{x,y} ({\mathbf{d}}) =  - \frac{D_m^2}{2\rho {\left( {2\pi
} \right)^3 } c_s} {\int\limits_{ - \infty }^\infty}
\frac{I_{x,y}(kd)}{\left( {1 + a_\textrm{B}^2 k^2 } \right)^4}
\frac{{c_s k^4dk}} {{c_s^2 k^2  - \Delta ^2}},
\end{equation}
can be evaluated along a contour in the upper complex plane. The
integration contour includes two simple poles, at
$k=\pm\Delta/c_s$, which have to be taken with weight $1/2$ due to
principal value integration. Also included is the pole at
$k=i/a_\textrm{B}$, of order four, and a simple pole at $k=0$
(with weight $1/2$). The latter pole is for the second term in
Eq.(\ref{eq:B:I-xy}) only.

The pole at $k=i/a_\textrm{B}$ yields exponentially decaying terms
$\sim \exp(-d/a_\textrm{B})$. At large $d$, the asymptotic
behavior is controlled by the poles at $k=\pm\Delta/c_s$,
\begin{equation}\label{eq:B:he-xy-c-asymptotic}
\chi _c^{x,y} ({\mathbf{d}}) \xrightarrow{{r\gg1}} - 4\pi ^2
C_\textrm{B} \frac{{2b}} {{\left( {1 + b^2 } \right)^4
}}\frac{{\sin br}} {{r^2 }},
\end{equation}
where $b = \Delta a_B /c_s$ and $r=|\mathbf{d}|/a_\textrm{B}$. The
complete expression can be easily obtained from
Eq.(\ref{eq:B:he-xy-c}). One can also note that at $d\rightarrow
0$, Eq.(\ref{eq:B:he-xy-c}) is
\begin{equation}\label{eq:B:he-xy-c-small-r}
\chi _c^{x,y} ({\mathbf{d}}) = - 4\pi ^2 C_B \frac{{1 + 9b^2  -
9b^4  + b^6 }} {{240\left( {1 + b^2 } \right)^4 }} + O(r^2 ).
\end{equation}

Along the same contour, the $z$ component of
Eq.(\ref{eq:S2:he-c}),
\begin{equation}\label{eq:B:he-z-c}
\chi _c^{z} ({\mathbf{d}}) =  - \frac{D_m^2}{2\rho {\left( {2\pi }
\right)^3 } c_s^2} {\int\limits_{ - \infty }^\infty}
\frac{I_{x,y}(kd)k^2dk}{\left( {1 + a_\textrm{B}^2 k^2 }
\right)^4}
\end{equation}
has only two poles, at $k=0$ [order 2 for the first term, and also
of order 1 or 3 for the second term in $I_z(kd)$], and at
$k=i/a_\textrm{B}$ (of order 4). The pole at $k=0$ is to be taken
with weight $1/2$ and gives the $1/r^5$ asymptotic,
\begin{equation}\label{eq:B:he-z-c-asymptotic}
\chi _c^{z} ({\mathbf{d}}) \xrightarrow{{r\gg1}} 384\pi ^2
C_\textrm{A} \frac{1} {{r^5 }},
\end{equation}
while for $r\rightarrow 0$ one obtains
\begin{equation}\label{eq:B:he-z-c-small-r}
\chi _c^z ({\mathbf{d}}) =  - C_A \frac{{\pi ^2 }} {{20}} +
O(r^2).
\end{equation}
Substituting Eqs.(\ref{eq:B:he-xy-c-asymptotic}) and
(\ref{eq:B:he-z-c-asymptotic}) in $H_\textrm{int}$, i.e., in the
second term in Eq.(\ref{eq:S2:H-eff}), one obtains
Eq.(\ref{eq:S5:H-int-asymptotic}).

By using Eqs.(\ref{eq:B:I-integral}) and (\ref{eq:B:I-xy}), for
Eq.(\ref{eq:S2:eta-c}) we obtain
\begin{equation}\label{eq:B:eta-xy-c}
\eta _c^{x,y} ({\mathbf{d}}) = \frac{\pi } {2}C_B \frac{{b^3 }}
{{\left( {1 + b^2 } \right)^4 }}I_{x,y} (rb)\coth \frac{\Delta }
{{2k_\textrm{B}T}}.
\end{equation}
The decoherence processes are dominated by the local noise terms,
e.g., $\eta_c^{x,y}(0)$. Noting that $I_{x,y}(0)=\frac{4\pi}{15}$,
one obtains (\ref{eq:S5:eta-c-xy}). One also finds that
$\eta_c^z(\mathbf{d})=0$. For low temperatures,
$\frac{\Delta}{2k_\textrm{B}T}\gg1$, one has $\coth\frac{\Delta
}{2k_\textrm{B}T}\approx 1$ and, therefore,
$\chi_s^{m}(\mathbf{d})\approx \eta_c^{m}(\mathbf{d})$, see
Eqs.(\ref{eq:S2:he-s}) and (\ref{eq:S2:eta-c}). Note that the
expressions involving $\chi_s^{m}(\mathbf{d})$ can be neglected in
(\ref{eq:S2:H-eff}), since the corrections they introduce to the
induced interaction have the same magnitude as the noise
amplitudes. The function $\eta_s^{m}(\mathbf{d})$ is often
comparable to $\eta_c^{m}(\mathbf{d})$ for Eq.(\ref{eq:S5:gk-3D})
with the parameters used in Section~\ref{Section6}. This amplitude
is calculated numerically in Figure~\ref{fig5}.


\end{document}